\documentstyle[aas2pp4]{article}
\def \ino         { \the\itemno\global\advance\itemno by 1 }

\def \kms       {\hbox{ km s$^{-1}$}}

\def \msol      {{\rm M}_\odot}

\def\kmpersec{km s$^{-1}$}

\lefthead{Governato et al.}
\righthead{Origin of early type galaxies}

\begin{document}

\title{On the origin of early type galaxies and the evolution of the
interaction rate in the field}

\author{F. Governato\altaffilmark{1,2},Jeffrey P. Gardner\altaffilmark{1}, 
J. Stadel\altaffilmark{1}, T. Quinn\altaffilmark{1} \& G. Lake\altaffilmark{1}}

\affil{Astronomy Department, University of Washington, Seattle, WA 98195}

\altaffiltext{2}{Present address:  Physics Department, Univ. of Durham, 
Science Labs, South Road, Durham, UK}

\begin{abstract}

  Using cosmological N-body simulations of critical (SCDM) and open
  ($\Omega$=0.3, OCDM) cold dark matter models we have identified dark
  matter halos which are associated with the progenitors of present
  day bright early--type galaxies.  By following their merging
  history, we show how early--type galaxies that formed within massive
  halos at redshift $\simeq 3$ are now preferentially residing in
  clusters and groups.  On the other hand, those that formed through
  major merging events between redshift 1 and the present have not yet
  been accreted into larger, virialized structures.  This result is in
  agreement with analytical predictions in hierarchical clustering
  models.  CDM models are able to explain both the ancient and uniform
  population of ellipticals that dominates in clusters together with
  the more recent and heterogeneous population of field ellipticals.
  Predictions for the comoving number density of bright early--type
  galaxies are given, and are shown to be consistent with the observed
  luminosity function.  We predict that the number density of
  interacting bright binary galaxies, from which the field population
  of ellipticals may have originated, is proportional to
  (1+z)$^{4.2\pm0.28}$ and (1+z)$^{2.5\pm0.42}$ in SCDM and OCDM
  respectively. This result is consistent with previous analytical
  estimates and is discussed together with recent observational
  constraints.

\end{abstract}


\section{Introduction}

Most elliptical galaxies,  are well described by the fundamental plane
and show the same dynamical properties (Djorgovski \& Davis 1987,
Faber {\it {et al.}} 1987, Guzman, Lucey \& Bower 1993).  However, one
may also find in the literature a number of publications, including
classic work by Larson, Tinsley, \& Caldwell (1980), which point out
that bright elliptical galaxies residing in clusters have major
differences in their stellar populations compared to those residing in
the field.  Cluster ellipticals show a surprisingly tight
color-magnitude and M$gb-\sigma$ relation both at the present (Larson
{\it {et al.}} 1980, Bower {\it {et al.}}  1992) and at higher $z$
(Ellis {it et al} 1997, Ziegler \& Bender 1997). Moreover their colors
are consistent with the bulk of their stellar population having been
formed at $z>2$ (Bender, Ziegler, \& Bruzual 1996; Ellis {\it et al.}
1997).  The population of field ellipticals shows instead a much
larger scatter, indicating a younger age as well as a spread in the
time of their last major starburst of at least a few Gyrs (De Carvalho
\& Djorgovski 1992; Rose {\it {et al.}} 1994; Longhetti {\it {et al.}}
1997).  They also possess a number of features (shells,
counterrotating inner disks), believed to be associated with their
origin by merger events, which correlate with bluer colors (Schweizer
\& Seitzer 1992). While deviation from pure ellipsoids, (i.e.  boxy or
disky isophotes) is not obviously linked to a recent merger event
(Governato, Reduzzi, \& Rampazzo 1993; Heyl, Hernquist, \& Spergel
1994), strongly disky isophotes are interpreted as a sign of recent
star formation in a small central gaseous disk, which could have
perhaps been accreted through a merger (de Jong \& Davies 1996).
Finally, Kauffmann, Charlot, \& White (1996) point out that a
consistent fraction of galaxies with red colors, possibly in the
field, have experienced star formation activity at redshifts $<1$.

These contrasting results have led to the ``nature'' or ``nurture''
hypotheses for the formation of early type galaxies, and specifically
of ellipticals (Es).  In the former case, star formation occurred at
high redshift in a rapid ($\sim 1$ Gyr) burst within protogalactic
halos which then quickly coalesced to form galaxies with a dominant
spheroidal component; after this event their stellar population
evolved passively.  In the latter instance, mergers between possibly
gas rich galaxies created spheroidal galaxies as remnants (Toomre \&
Toomre 1972).  These two models, can be considered two extreme methods
of forming Es within the more general hierarchical clustering
framework, where dark matter (DM) halos merge continuously through
gravitational instabilities, creating larger and larger structures.
This scenario has received strong support from both semianalytical
(Baugh, Cole, \& Frenk 1996a; Kauffman 1996; Mo \& Fukugita 1996) and
numerical (Hernquist \& Barnes 1991; Hernquist 1993) work. In
particular, numerical results show that mergers between gas-rich disk
galaxies create systems with luminosity profiles, core structure and
kinematics similar to those observed in elliptical galaxies.

By making reasonable assumptions about the DM halos which host the
formation of early--type galaxies through the processes just
described, we explain the observed environment dependance in the
properties of elliptical galaxies within the cold dark matter (CDM)
scenario.

\section{The simulation dataset}

Our simulations followed the evolution of a periodic cube 100 Mpc on a
side in both a critical universe (SCDM) ($\Omega_0=1$, $h\equiv
H_0/100$ \kmpersec\ Mpc$^{-1}=0.5$, $\sigma_8 =0.7$) and an open
(OCDM) universe ($\Omega_0=0.3$, $h=0.75$, $\sigma_8 =1$).  The
normalizations were chosen to roughly match the observed cluster
abundance.  Each simulation was performed using PKDGRAV (Dikaiakos \&
Stadel 1996, Quinn, Katz, Stadel \& Lake 1998), a parallel N--body
treecode supporting periodic boundary conditions, and employed $144^3$
($\sim$ 3 million) particles with spline softening set to 60 kpc,
allowing us to resolve individual halos with present-day circular
velocities $V_c$ as low as 100 \kms with several tens of particles.
(In this paper  $V_c$ is derived directly from the mass of a given halo,
as obtained from the halo finder).
Halos were identified at each output using both a standard
friends-of-friends method and SKID, a halo-finding algorithm which
utilizes local density maxima (a copy can be obtained at 
http://www-hpcc.astro.washington.edu/tools/).

\section{The formation of early--type  galaxies}

As our simulations include only the effects of gravity on
collisionless matter, it is not possible to directly infer the
morphological types of galaxies that are thought to form inside DM
halos (White \& Rees 1978).  However, simple considerations based on
the expected the high accretion and merging rate , short dynamical
friction time-scales and the dynamics of gas cooling at high redshift,
allow us to safely assume that rare, massive halos formed at high
redshift will not be able to support a stable stellar disk similar to
those observed in present day spirals.  In fact analytical
considerations (Lacey \& Cole 1994, Bower 1991) suggest that major
mergers will be very common for galaxy sized halos at high redshift,
i.e. close to their typical formation time.  These halos will then
have a high chance to harbor galaxies with a dominant spheroidal
stellar component (Fall \& Efstathiou 1980; Toth \& Ostriker 1992;
Lacey \& Cole 1994; White 1997).  If those halos are subsequently
accreted into groups or clusters, the resultant large potential well
prohibits these galaxies from acquiring a gaseous disk through
secondary infall of cold gas.  Further support is given to this
scenario by recent observations of massively star forming galaxies at
high $z$ (Yee {\it et al.} 1996), as well as numerical simulations of
merging disk galaxies similar to present day spirals. The simulations
show that mergers with a mass ratio of 3:1 or less produce a remnant
resembling an elliptical galaxy (Barnes 1996).  If the merging event
happens at $z<1$ the galaxy does not have sufficient time to accrete a
new gaseous disk (Baugh {\it et al.}  1996b).  Also, at large absolute
magnitudes (L$>L_{*}$, i.e. large circular velocities) late type
galaxies become progressively rarer compared to early type ones (Lake
1990, Heyl {\it et al} 1996), so strengthening our assumption that
halos within the chosen mass range host a luminous early-type galaxy.

Based on these assumptions, we identified the halos in our simulations
which were most likely to host early-type galaxies that formed in each
of the two distinct ways suggested by the ``nature or nurture''
picture: i.e. massive halos formed at high redshift ($z >3$) by the
collapse of rare massive peaks and those that formed at $z<1$ by a
``major merger'' event, which was defined as the coalescing of two
distinct halos with a mass ratio of 3:1 or less.  The high-z halos
were required to have V$_c > 250 \kms$, similar to what
observed for many high redshift star forming galaxies (Steidel {\it el al.}
1996) while the range used to identify the low-z merger remnants was
220 \kms$ \leq V_c \leq 320$ \kms, typical of bright present day
Es.  Our results do not depend on the details of the velocity
cutoffs used, which have been chosen to be representative of those
observed in real objects.

Moreover we assume that within our chosen range, the $V_c$ of a given
halo can be associated with the $V_c$ of the main galaxy resident
within that halo.  Galaxies whose stellar populations formed at high
redshift ( $\sim 2$ or higher) will show very uniform ages, as they
all formed when the age of the Universe was just a small fraction of
the present. On the contrary a sample of galaxies that formed a
consistent part of their star content at $z<1$, will show a larger
spread in colors and derived ages.

It is interesting to note that theoretical models (Baugh {\it et al}
1997, Governato {\it et al} 1997) of galaxy formation predict that
Lyman break galaxies like those identified at $z\sim 3$ by Steidel and
collaborators (1996) will preferentially be residing in halos with
circular velocity close to our range of choice.  Most likely, the
selection criteria used in our work do not define the entire class of
halos that could host the formation of early type galaxies and Es in
particular; rather, these two populations delimit the extreme ways of
forming early-type galaxies within the hierarchical assemblage of
cosmic structures.  For example S0s could have originated from a
variety of different merging histories, such as multiple small
accretions of small satellites or fast two body encounters inside a
group or cluster environment (Miller 1983, Moore {\it et al.} 1996;
Oemler, Dressler, \& Butcher 1997) that lead to short, intense
starbursts (Poggianti \& Barbaro 1997) and consumed their gas content.

To study the evolution of the population of early type galaxies
associated with the identified halos, we examined their subsequent
evolution, as made possible by the numerical simulation. To do so we
selected halos that, at the present time,  contain  the halos selected
at higher redshift.

We used SKID to identify both high and low redshift halo populations
because of its ability over the friends-of-friends algorithm (FOF) to
isolate distinct halos even in overdense regions and in the process of
merging.  Moreover, FOF tends to link together binary halos which
would compromise our study. Instead, halos at $z=0$ were identified
using FOF, since we were no longer interested in identifying
substructure within larger halos, but rather in defining halo masses
within a canonical overdensity $\delta\rho/\rho \sim 178$ relative to
the critical density, corresponding to the boundaries of virialized
structures.  Our results do not change significantly if just one or
the other of the two halo-finding algorithms are used, although using
solely FOF tends to overestimate the redshift of major mergers since
the interacting binaries are ``linked together'' into a single object
sooner than with SKID.  FOF also finds fewer high-z objects as it
tends to link individual objects that are close to each other and
inside a common overdense region.

Theoretical models of clustering (Bardeen {\it et al.} 1986) predict
that rare peaks at one mass scale, as the halos we selected, should be
highly correlated in space.  It is interesting to show how well this
statement holds in a N-body experiment for actual halos likely to
be associated with early-type galaxies, since these will be more
closely related to observational data.

\begin{figure}
\plotone{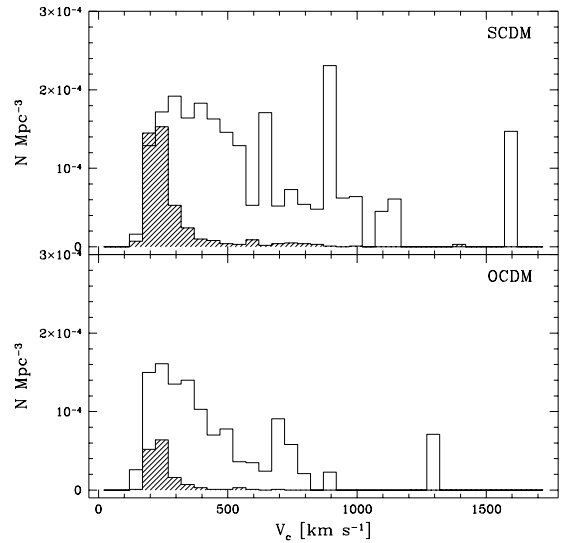}
\caption{ Histogram of the comoving number density distribution of
 early--type galaxies identified at $z=3$ as a function of the
 circular velocity of halos that contain them at the present time.
 The top panel shows results for  SCDM  and the bottom panel
 from OCDM.  The last bin represents the biggest halo identified at
 $z=0$, which contains several tens of halos identified at $z=3$.  The
 continuous line refers to halos selected at $z=3$, with $250 \kms
 \leq V_c$.  The shaded region refers to halos formed by major mergers
 between $1>z>0$ with circular velocity 220 \kms$ \leq V_c \leq 320$.
}
\label{Fig.1}
\end{figure}

\begin{figure}
\plotone{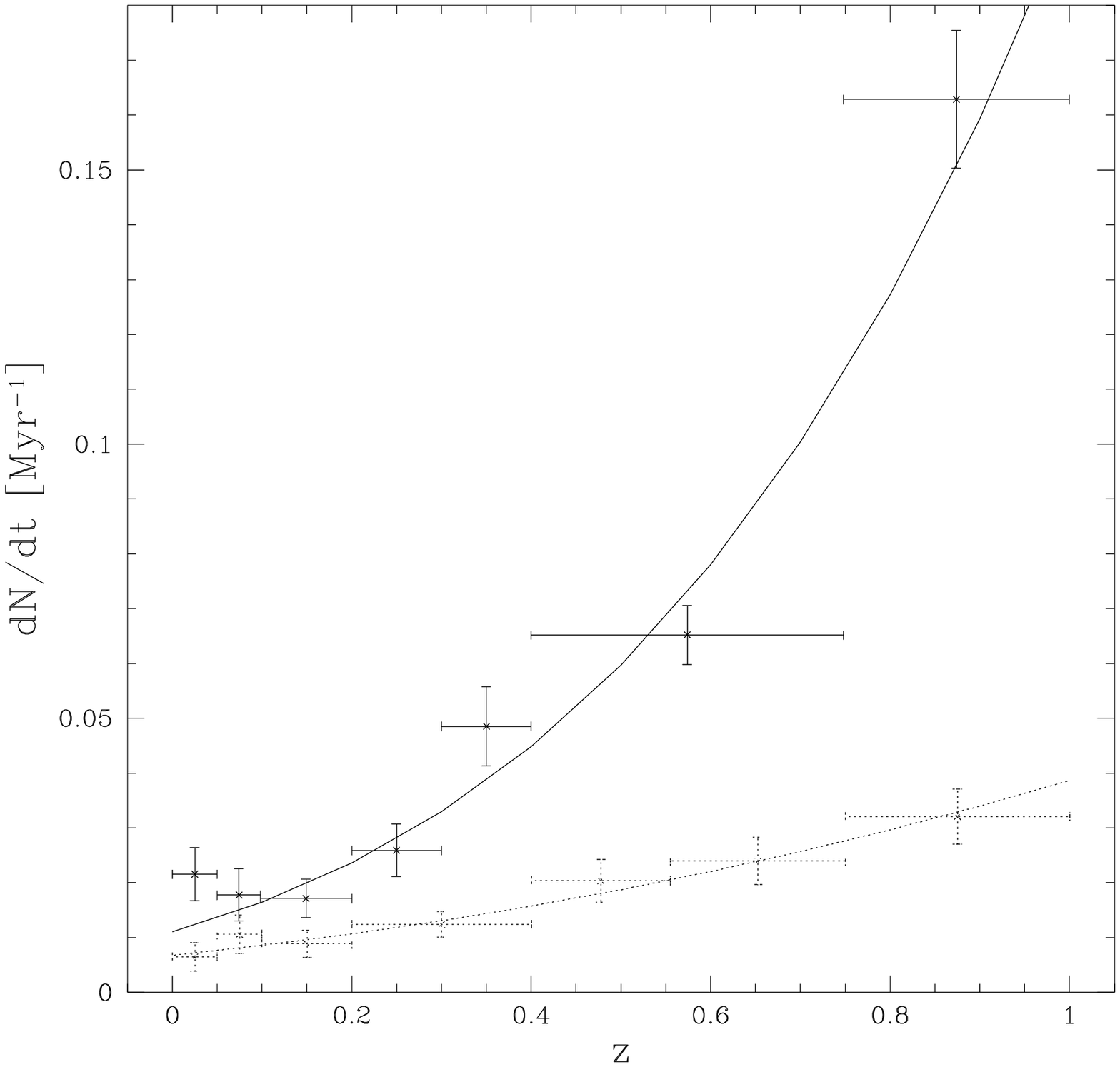}
\caption{Rate of formation of major mergers between $0<z<1$ as a
  function of redshift for SCDM (solid) and OCDM (dotted). Lines are
  best fits in the form (1+z)$^\alpha$, where $\alpha=4.2$ for SCDM
  and 2.5 for OCDM.
}
\label{Fig.2}
\end{figure}

It is evident from Fig.1 that for both cosmologies, the majority of
halos within our velocity range selected at $z=3$ or higher now reside
in halos with large circular velocities, comparable to groups or
galaxy clusters. Visually examining the simulations we find that the
halos identified at $z=3$ are strongly correlated and trace the
forming large scale structure. Based on the variance in halo number
counts, the total halo population is more clustered than the mass, (by
a factor of 2 at a scale of 8$h^{-1}$Mpc in the SCDM run).  This bias
is even higher ($b\sim 4$) for the most massive halos, as those we
identified at $z$ = 3.  They are preferentially located along the
biggest filaments (Sathyaprakash, Sahni, \& Shandarin 1996). Peculiar
motions are preferentially along the filaments, as halos move towards
filament intersections were groups and clusters are just beginning to
form. This accretion of galaxies, already formed at rather high
redshift, continues for a long period of time, even to the present in
the critical model. In fact, most of groups and clusters are typically
assembled as a single halo at redshift below one.

Our findings for the $z=3$ selected halos do not imply that the
original galaxies have necessarily merged into a single central galaxy
by the present time. Tidal stripping and the extremely long time scale
for dynamical friction decay (of the order of the Hubble time) should
prevent a large fraction of the galaxies from reaching the center of
the cluster once it has been stripped of its DM halo. This hypothesis
is also supported by recent high resolution simulations of cluster
formation (Ghigna {\it et al} 1997).  At the same time, however, the
deep cluster potential well would prevent these galaxies from
acquiring a substantial gaseous component to fuel star formation at
later times.  Consequently, we argue that these galaxies are likely be
the progenitors of the observed population of old cluster Es.
 
Recent results (Dickinson 1997, Oemler {\it et al.} 1997, Smail {\it
 et al} 1997) show that even in clusters at $z\geq 0.5$ the elliptical
 galaxy population is already in place, showing colors typical of
 passive evolution and consistent with an origin at z$>2$, and so
 supporting our proposed scenario. Even if the accretion of
 Es by clusters happened over a large range of z, in parallel
 with the build up of clusters from smaller structures, this did not
 create a population of elliptical galaxies in clusters with a large
 age spread, suggesting that the initial collapse of the parent halos
 was efficient in turning most of each individual galaxy gas content
 into stars or expelling it under the form of hot gas.

On the other hand, a significant fraction of the S0 galaxies observed
at the present time seems to be lacking in clusters at intermediate
$z$, suggesting a different time scale and principle mechanism for
their creation, possibly due to interactions within the cluster
environment.

The situation is rather different if we look at the products of major
mergers selected between $1>z>0$. They now reside in halos of mass
comparable to that at their time of formation, indicating that they
have not yet been accreted by a larger structure.  The great majority
of them reside in what is generically referred to as ``the field'',
i.e regions of density close to the average one, implying that merger
remnants are much less biased compared to the general mass
distribution.  Even these halos also formed within a very complex
environment made of filaments and planes; however gravitational
clustering did not have enough time to act and move them into 
larger structures, as was possible for halos selected at $z = 3$.

It is interesting to compare the number density of our selected halos
with existing observational data. The luminosity function obtained by
Heyl {\it et al.} (1996) gives a number density of $2\times 10^{-3}$
Mpc${^{-3}}$ for ``red E'' galaxies with $M_{\hbox{bj}}<-19$, the
magnitude cut corresponding to the circular velocity associated to our
lower limit.  The Marzke {\it et al.} (1994) luminosity function
(based on Zwicky's magnitudes) gives a comoving number density of E+S0
in the same luminosity range of $1.8\times 10^{-3}$Mpc${^{-3}}$.  If
we sum the halos selected at $z=3$ with all major mergers between
$1>z>0$ we obtain a number density of $2.8\times 10^{-3}$ Mpc${^{-3}}$
and $1.4 \times 10^{-3}$ Mpc$^{-3}$ for SCDM and OCDM respectively.
In both cosmologies early--type galaxies identified at high $z$ are
much more numerous than those that were formed at lower redshift. In our
scenario recent merger remnants would represent only 16\% (for SCDM)
and 10\% (for OCDM) of the present day population of elliptical
galaxies.

Considering the large uncertainties in both our schematic models and
the observational data (due for example, to the size of the volume
studied and hence of the small range of environments sampled) there is
a good agreement, within less than a factor of two, between theory and
the observed numbers.  A better match could be found for individual
models but this would require to arbitrarily adjust the velocity
cutoff we used to select halos and/or the magnitude cut in the
luminosity function at the present time.  Overall, though, the
agreement between the observed number density of Es and the number density
of the selected halos in the simulations strongly supports the
hypothesis that we are selecting DM halos associated with galaxies
dominated by a spheroidal component.

These results illustrate that two different and possibly extreme
populations of halos that should host early-type galaxies are indeed
very segregated in space, consistent with both analytical predictions
for the clustering of halos and with observational evidence for Es.
In this scenario, cluster and field Es would naturally show different
age properties.  Galaxies that most probably formed their stellar
population at high redshift ($z>2$) would have uniform colors,
consistent with a similar age of formation followed by passive
evolution. Es formed by late mergers would instead have a larger
spread in the age, corresponding to their latest major starburst.  In
agreement with observations, only a few of this galaxies now reside in
groups and clusters. Hence, CDM models provide a consistent framework
capable of explaining the differences in age and properties between
the bulk populations of cluster Es (ancient and with uniform age for
their stellar component ) and field Es (younger and with a larger age
spread).

\section {The interaction rate of binary systems}

There is strong observational evidence (Driver {\it et al.} 1995,
Glazebrook {\it et al} 1995) that the number of interacting systems
grows rapidly with look-back time.  New data from the Hubble Deep
Field (Abraham {\it et al.} 1996; Van Den Bergh {\it et al.} 1996) and
recent redshift surveys (Patton {\it et al} 1996) make it possible to
measure their number and determine whether the Universe was in the
past more dynamically active at galactic scales.  The data suggest a
strong evolution in number density proportional to $(1+z)^{2.8\pm0.9}$
(Patton {\it et al.} 1996) or even steeper: ($\propto
(1+z)^{4.0\pm2.5}$ in Zepf \& Koo (1989)) or $\propto
(1+z)^{3.4\pm1.0}$ as suggested by the number of close galaxy pairs
(Carlberg, Pritchet \& Infante 1994).

 The exact value for this trend is very important, given its
consequences on the evolution of the galaxy population and on galaxy
counts (Ellis 1997, Roche {\it et al.}  1996, Baugh {\it et al.}
1996b).  Similar trends are found in samples of radio-quiet QSOs
(Boyle {\it et al.} 1993) and IRAS-selected galaxies.  Fig.2 plots the
rate of formation of merger remnants for SCDM and OCDM models.
Vertical error bars represent statistical uncertainties due to the
finite size of the sample, and the horizontal span the time interval
between successive outputs (of the order of one Gyr), effectively our
bin size.  We find (1+z)$^{4.2\pm0.28}$ for SCDM and
$(1+z)^{2.5\pm0.42}$ for OCDM respectively, while the rate at $z=0$ is
similar and close to $10^{-8}$Myr$^{-1}$Mpc$^{-3}$.  Merging rates
obtained from this work are in good agreement with previous analytical
predictions by Carlberg (1990).  Our selected sample of halos should
host luminous galaxies, which in turn should be preferentially
selected in samples containing higher redshift galaxies. As suggested
by detailed numerical simulations, they should form a luminous
starburst remnant when the galaxies inside each halo actually collide
and merge (Mihos \& Hernquist 1994).  While taken at face value this
result would seem to slightly favor a low omega model, were the
interaction rate is lower and the redshift dependency flatter, it is
 difficult to translate our results into a strong number
density prediction for a specific class of observed objects.  For
example, Lavery {\it et al.}(1996) suggest a rate of interactions of
$(1+z)^{4.5}$ based on the number density of ring galaxies.  In fact
it is not yet clear how QSOs and starburst galaxies, of which the
number density is found to increase with redshift, are directly
related to interactions and mergers.  However our results suggest a
general agreement of CDM models with the observed trend.  It is
interesting to note that, due to the steeper increase at high
redshift, and contrary to what would be naively expected, the mean
redshifts for this specific class of mergers is higher in the SCDM
sample than in the OCDM one. However, the longer age stretch in the
OCDM universe compensates for this, once measured in look back time.

In the hierarchical clustering model, collapse begins with smaller
(i.e. galaxy-sized) scales and proceeds to larger and larger masses.
Hence, the observed increase in galactic activity in the past is a
natural by-product of this model, with the number of interacting
galaxies increasing with redshift.  In the present-day universe, most
of the galaxy-sized perturbations have collapsed, and it is now
cluster-scale objects with masses of order $10^{14}\msol$ or higher
that are being assembled.  Clusters are indeed observed to be forming
and accreting large quantities of matter at the present time, but
there is no evidence of an increase in their formation rate looking at
high redshifts (Eke, Cole, \& Frenk 1996).  The ability to predict
this very diverse, time-dependent behavior between objects at
different mass scales can be considered a major success of the
hierarchical clustering model and specifically of CDM.

\section {Discussion}

In this paper we show how hierarchical clustering models, namely CDM,
naturally account for the the different properties of cluster and
field Es. In particular they account for the high average value and
the small scatter in the ages measured for cluster Es as opposed to
the younger age as well as a larger scatter around the average value
found for Es in the generic field environment.

Numerical simulations show, in accordance with theoretical
predictions, that spatial correlation of the higher peaks in the
density field (Bardeen {\it et al.} 1986) naturally explains  the
fact that the objects that collapse at high redshift are now very
clustered and  contributed to the build up of galaxy clusters.
Instead, galaxy sized halos collapsing at later times show a much less
degree of clustering and are preferentially found in field regions.

 In fact, after making the hypothesis that stellar population ages are
tied to the collapse times of the parent DM halos, we are able to show
that early-type galaxies that formed their stellar population at high
redshift in massive DM halos are now in clusters, while those formed
through major mergers at $z<1$ have not been accreted into larger
structures. As these recently formed galaxies do not end up into dense
environment, elliptical galaxies in clusters should show, for the
great majority, colors consistent with a very ancient stellar
population.  This is consistent both with observations and with
predictions from semianalytical models of galaxy formation (Kauffman
1996). The small scatter in age for cluster Es arises naturally from
the fact that at high z, when they formed the majority of their
stellar content, the age of the Universe was a small fraction of the
present time.  Apparently, even if these galaxies were accreted into
cluster at different times, subsequent to their formation, this
delayed accretion did not create a large scatter in the age of their
stellar populations.

We further demonstrate that within the CDM framework the interaction
rate in binary galactic systems increases rapidly in the past,
consistent with observations.  Recent techniques based on line
strength indices are able to evaluate the epoch of the last starburst
with a precision of a few Gyrs (Dorman \& O'Connel, 1996; Bressan,
Chiosi, \& Tantalo 1996; De Jong \& Davies 1997; Longhetti {\it et
al.} 1997) and disentangle the time evolution from metal abundances of
the stellar populations.  These methods will prove invaluable in
tracing the origin of early type galaxies in different environments
and will provide a larger database to test theories of galaxy
formation.  Future detailed simulations which include hydrodynamics
and star formation processes will be able to make more robust and
quantitative predictions about the origin and evolution of galaxies
within the CDM scenario.

\acknowledgments

We thank Richard Bower, Sandra Savaglio, Ian Smail and Bianca Maria
 Poggianti for useful comments and discussions.  We acknowledge
 extensive use of the ADS and astro-ph preprint databases.
 Simulations were done at the ARSC and NCSA computing centers. Songs
 by the Rolling Stones were helpful during the writing of this paper.

%
%

\end{document}